\documentclass[aps,prd,onecolumn,superscriptaddress,showpacs,nofootinbib,floatfix,onecolumn]{revtex4}
\usepackage{dcolumn}
\usepackage{bm}

\usepackage[dvips]{graphicx}
\usepackage{float}
\usepackage{epsfig}
\usepackage{graphicx}
\usepackage{longtable}
\usepackage{rotating}
\usepackage{ulem}
\usepackage{subfigure}
\usepackage{amsmath,amssymb}
\usepackage[colorlinks=true,allcolors=blue]{hyperref}

\def\ttt#1{\texttt{\small #1}}

\providecommand{\ee}{e$^+$e$^-$}
\providecommand{\pp}{p-p}

\providecommand{\pPb}{p-Pb}

\providecommand{\PbPb}{Pb-Pb}
\providecommand{\gaga}{\gamma\,\gamma}

\providecommand{\mgg}{\rm m_{\gamma\gamma}}
\providecommand{\sigmagg}{\sigma_{\gaga\to\gaga}}

\newcommand{\sgg}{s_{_{\gamma\,\gamma }}}
\newcommand{\sqrtsgg}{\sqrt{s_{_{\gamma\,\gamma }}}}
\newcommand{\sqrtsnn}{\sqrt{s_{_{\rm NN}}}}
\newcommand{\qqbar}    {\ensuremath{q\bar{q}}}

\newcommand{\ABgaga}{A\,B\,$\xrightarrow{\gaga}$ A\,$\gaga$\,B}
\newcommand{\ABglgl}{A\,B\,$\xrightarrow{g\,g}$ A\,$\gaga$\,B}

\providecommand{\madgraph}{{\sc MadGraph}}
\providecommand{\superchic}{{\sc SuperChic}}

\newcommand{\Lumi}{\mathcal{L}}

\newcommand{\Pom} {I\!P} 

\begin{document}

\title{Observing light-by-light scattering at the Large Hadron Collider}
 
\author{David~d'Enterria}
\affiliation{CERN, PH Department, 1211 Geneva, Switzerland}

\author{Gustavo G.~Silveira}
\affiliation{UC Louvain, Center for Particle Physics and Phenomenology (CP3), Louvain-la-Neuve, Belgium}

\begin{abstract}
\noindent
Elastic light-by-light scattering ($\gaga\to\gaga$) is open to study at the Large Hadron Collider thanks to 
the large quasi-real photon fluxes available in electromagnetic interactions of protons (p) and lead (Pb) ions.
The $\gaga~\to~\gaga$ cross sections for diphoton masses $\mgg>$~5~GeV amount to 105~fb, 260~pb, and 370~nb 
in \pp, \pPb, and \PbPb\ collisions at nucleon-nucleon center-of-mass energies $\sqrtsnn$~=~14 TeV, 8.8~TeV,
and 5.5~TeV respectively. Such a measurement has no substantial backgrounds in 
\PbPb\ collisions where one expects about 70 signal events per run,
after typical detector acceptance and reconstruction efficiency selections.
\end{abstract}


\maketitle

\begin{figure}[hbt!]
\centering
\includegraphics[height=4.5cm]{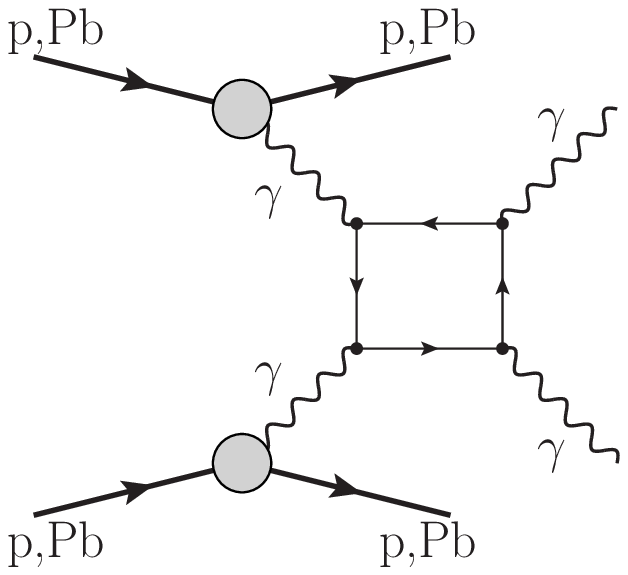}
\caption{Schematic diagram of elastic $\gaga\to\gaga$ collisions in electromagnetic proton and/or ion interactions at the LHC.
 The initial-state photons are emitted coherently by the protons and/or nuclei which survive the electromagnetic interaction.} 
  \label{fig:diag}
\end{figure}

\paragraph*{Introduction.  --}
The elastic scattering of two photons in vacuum ($\gaga\to\gaga$) is a pure quantum-mechanical process
that proceeds at leading order in the fine structure constant, $\mathcal{O}(\alpha^4)$, 
via virtual one-loop box diagrams containing charged particles (Fig.~\ref{fig:diag}).
Although light-by-light (LbyL) scattering via an electron loop has been precisely, albeit indirectly, tested in the 
measurements of the anomalous magnetic moment of the electron~\cite{VanDyck:1987ay} and muon~\cite{Brown:2001mga},
its direct observation in the laboratory remains elusive still today. Out of the two closely-related processes
--photon scattering in the Coulomb field of a nucleus (Delbr\"uck scattering)~\cite{Milstein:1994zz} and
photon-splitting in a strong magnetic field (``vacuum'' birefringence)~\cite{Adler:1971wn,Bregant:2008yb}-- 
only the former has been clearly observed~\cite{Jarlskog:1974tx}. 
Several experimental approaches have been proposed to directly detect $\gaga\to\gaga$ in the laboratory
using e.g.~Compton-backscattered photons against laser photons~\cite{Mikaelian:1981fh}, collisions of photons 
from  microwave waveguides or cavities~\cite{Brodin:2001zz} or high-power lasers~\cite{Moulin:1996vv,Lundstrom:2005za}, 
as well as at photon colliders~\cite{Brodsky:1994nf,Jikia:1993tc} where energetic photon beams can be obtained by 
Compton-backscattering laser-light off electron-positron (\ee) beams~\cite{Ginzburg:1981ik}. Despite its fundamental
simplicity, no observation of the process exists so far.\\

In the present letter we investigate the novel possibility to detect elastic photon-photon scattering using the large
(quasi-real) photon fluxes of the protons and ions accelerated at TeV energies at the CERN Large Hadron Collider (LHC). 
In the standard model (SM), the box diagram depicted in Fig.~\ref{fig:diag} involves charged fermions (leptons
and quarks) and boson (W$^\pm$) loops. 
In extensions of the SM, extra virtual contributions
from new heavy charged particles are also possible. 
The study of the $\gaga\to\gaga$ process --in particular at the high invariant masses reachable at photon
colliders-- has thus been proposed as a particularly neat channel to study anomalous 
gauge-couplings~\cite{Brodsky:1994nf,Jikia:1993tc}, new possible contributions from charged supersymmetric
partners of SM particles~\cite{Ohnemus:1993qw}, monopoles~\cite{Ginzburg:1998vb}, and
unparticles~\cite{Cakir:2007xb}, as well as low-scale gravity effects~\cite{Cheung:1999ja,Davoudiasl:1999di}
and non-commutative interactions~\cite{Hewett:2000zp}. 

Photon-photon collisions in ``ultraperipheral'' collisions 
of proton~\cite{d'Enterria:2008sh,louvain} and lead (Pb) beams~\cite{Baltz:2007kq} have been
experimentally observed at the LHC~\cite{Chatrchyan:2011ci,Chatrchyan:2012tv,Chatrchyan:2013foa,Abelev:2012ba,Abbas:2013oua}.
All charges accelerated at high energies generate electromagnetic fields which, in the equivalent 
photon approximation (EPA)~\cite{WW}, can be considered as $\gamma$ beams~\cite{Stan70}.
The emitted photons are almost on mass shell, with virtuality $- Q^{2} < 1/R^{2}$, where $R$ 
is the radius of the charge, i.e. $Q^2\approx$~0.08~GeV$^2$ for protons with $R\approx$~0.7~fm,
and $Q^2<$~4$\cdot10^{-3}$~GeV$^2$ for nuclei with $R_{\rm A}\approx 1.2\,A^{1/3}$~fm, for mass number $A>$~16. 
Naively, the photon-photon luminosities are suppressed by a factor
$\alpha^2\approx$~5$\cdot$10$^{-5}$ and only moderately enhanced by logarithmic corrections $\propto\ln^3(\sqrtsnn)$,
compared to the corresponding hadronic beam luminosities. However, since each photon flux scales 
as the squared charge of the beam, $Z^2$, $\gaga$ luminosities are extremely enhanced for ion beams,
up to $Z^4$~=~5$\cdot$10$^{7}$ in the case of \PbPb\ collisions. 
The photon spectra have a typical $E_{\gamma}^{-1}$ power-law fall-off up to
energies of the order of 
$\omega_{\rm max}\approx\gamma/R$, where $\gamma=\sqrtsnn/(2$m$_{\rm N})$ 
is the Lorentz relativistic factor of the
proton (m$_{\rm N}$~=~0.9383~GeV) or ion (with nucleon mass m$_{\rm N}$~=~0.9315~GeV), 
beyond which the photon flux is exponentially suppressed. 
Although the $\gamma$ spectrum is harder for smaller charges --which favors proton over nuclear beams in the
production of diphoton systems with large invariant masses-- 
the $\gaga\to\gaga$ cross section decreases rapidly, as the square of the center-of-mass (c.m.) energy
$\sim\sgg^{-1}$ from its peak at $\sqrtsgg\approx$~3m$_{\rm e}$~\cite{Bern:2001dg}, which favors the
comparatively softer Pb photon beams for the observation of LbyL scattering.
In Table~\ref{tab:1} we summarize the most relevant parameters for ultraperipheral \pp, \pPb, and \PbPb\ 
collisions at the LHC~\cite{Baltz:2007kq,Salgado:2011wc}.\\ 

The final-state signature of interest here is the exclusive production of two photons,
\ABgaga\ where the diphoton final-state is measured in the central detector, and
the incoming hadrons A,B~=~p,Pb survive the electromagnetic interaction and are
scattered at very low angles with respect to the beam.
The very same final-state can be mediated by the strong interaction through a quark-loop in 
the exchange of two gluons in a color-singlet state, \ABglgl~\cite{Khoze:2004ak}. Such 
``central exclusive production'' (CEP), observed in p$\bar{\rm p}$ collisions at 
Tevatron~\cite{Aaltonen:2011hi} and searched for at the LHC~\cite{Chatrchyan:2012tv},
constitutes an important background for the $\gaga\to\gaga$ measurement in \pp\ but not for 
\PbPb\ collisions as discussed later.\\ 


\paragraph*{Theoretical setup.  --}

The elastic $\gaga$ production cross section via photon-photon fusion in the collision of hadrons
A and B factorizes into the product of the elementary cross section for $\gaga\rightarrow \gaga$ at
 $\sqrtsgg$, convoluted with the EPA spectra from the two colliding beams:
\begin{equation}
\sigmagg^{\rm excl}=\sigma(\rm{A}\; \rm{B}\,\xrightarrow{\gaga} \rm{A} \; \gaga \; \rm{B})=
\int d\omega_1 d\omega_2 \, \frac{f_{\rm \gamma/A}(\omega_1)}{\omega_1}\, \frac{f_{\rm \gamma/B}(\omega_2)}{\omega_2}\; \sigmagg(\sqrtsgg),
\label{eq:two-photon}
\end{equation}
where $\omega_1$ and $\omega_2$ are the two photon energies, and $f_{\rm A,B}(\omega)$ 
are the photon fluxes at energy $\omega$ emitted by the hadrons A and B. 
The photon energies determine the rapidity $y$ of the produced system $y = 0.5\, \ln(\omega_1/\omega_2)$
and the c.m.{} energy $\sqrtsgg=\mgg=\sqrt{4\omega_1\omega_2}$ which,
for symmetric systems, is maximal at $y=0$ when 
$\omega_1^{\rm max}=\omega_2^{\rm max} \approx \gamma/b_{\rm min}$ 
with $b_{\rm min}$ the minimum separation between the two charges of radius $R_{\rm A,B}$.
We use the proton $\gamma$ spectrum obtained from its elastic form factor~\cite{Budnev:1974de} 
and, for the ion $\gamma$ spectrum, the impact-parameter dependent expression
integrated from $b_{\rm min}$ to infinity~\cite{Bertulani:1987tz} with the requirement 
$b_{\rm min}=R_{\rm A,B}$ plus a correction equivalent to the geometrical condition 
$|\vec b_{\rm 1} - \vec b_{\rm 2}| > R_{\rm A}+R_{\rm B}$~\cite{Cahn:1990jk} to ensure 
that all collisions occur 
without hadronic overlap and breakup of the colliding beams. 
Propagated uncertainties to the final cross sections are of order 
$\pm$10\% ($\pm$20\%) for \pp\ and \pPb\ (\PbPb) collisions, covering different 
form-factors parametrizations and the convolution of the nuclear photon fluxes.\\

We use the \madgraph\ v.5 Monte Carlo (MC)~\cite{madgraph} framework to implement the elastic p and 
Pb photon fluxes, and the leading-order expression for the $\sigmagg$ cross section~\cite{Bern:2001dg}
including all quark and lepton loops. 
We omit the W$^\pm$ contributions which are only important for diphoton masses $\mgg\gtrsim$~2\,m$_{\rm W}$.
Inclusion of next-to-leading-order QCD and QED corrections increases $\sigmagg$ by a few
percent~\cite{Bern:2001dg}, but taking into account a gap survival factor of $\hat{S}^2$~=~0.9--1.0 
--encoding the probability to produce fully exclusively the diphoton system without 
any other hadronic activity from soft rescatterings between the colliding hadrons~\cite{Khoze:2004ak}-- would
reduce the yields by about the same amount.\\ 

\begin{table}[htpb]
\begin{center}
\caption[]{Parameters for the $\gaga\to\gaga$ measurement in A\,B collisions at the LHC:
(i) nucleon-nucleon c.m.{} energy, $\sqrtsnn$, (ii) integrated luminosity $\Lumi_{\rm AB}\cdot\Delta t$ 
(where $\Lumi_{\rm AB}$ are the beam luminosities --for low pileup in the \pp\ case, see text-- and a year
is defined as $\Delta t$~=~10$^{7}$~s for \pp, and 10$^{6}$~s for \pPb\ and \PbPb), 
(iii) beam Lorentz factor, $\gamma$, 
(iv) effective radius of the (largest) charge, $R_{\rm A}$, 
(v) photon energy tail in the c.m.{} frame, $\omega_{\rm max}$, 
(vi) maximum photon-photon c.m.{} energy, $\sqrt{s_{\gaga}^{\rm max}}$, (vii) exclusive $\gaga\to\gaga$ cross section for
diphoton masses above 5~GeV, and (viii) expected number of signal counts per year after selection cuts (see text).}
\label{tab:1}
\vspace{0.4cm}
\begin{tabular}{l|cccccc|cc} \hline 
System  & \hspace{0.1cm} $\sqrtsnn$ \hspace{0.1cm}& \hspace{0.1cm}${\cal L}_{\rm AB}\cdot\Delta t$ \hspace{0.1cm} &
\hspace{0.1cm} $\gamma$ \hspace{0.1cm} &
\hspace{0.1cm} $R_{\rm A}$ \hspace{0.1cm} & \hspace{0.1cm} $\omega_{\rm max}$ \hspace{0.1cm} 
& \hspace{0.1cm} $\sqrt{s_{\gaga}^{\rm max}}$ \hspace{0.1cm} 
& \hspace{0.1cm} $\sigmagg^{\rm excl}$\hspace{0.1cm} 
& \hspace{0.1cm} $N_{\gaga}^{\rm excl}$ (per year) \hspace{0.1cm} \\
       & (TeV) &  (per year) &  & (fm) & (GeV) & (GeV) &  \hspace{0.2cm} [$\mgg>$~5~GeV] &  \hspace{0.2cm} [m$_{\gaga}>$~5~GeV, after cuts] \\ \hline
\pp    & 14  & 1~fb$^{-1}$   & 7455 & 0.7 & 2450 & 4500 & 105 $\pm$ 10 fb & 12 \\ 
\pPb   & 8.8 & 200~nb$^{-1}$ & 4690 & 7.1 & 130  & 260  & 260 $\pm$ 26 pb & 6 \\ 
\PbPb  & 5.5 & 1~nb$^{-1}$   & 2930 & 7.1 &  80  & 160  & 370 $\pm$ 70 nb & 70 \\ \hline
\end{tabular}
\end{center}
\end{table}


\paragraph*{Results. --}

\begin{figure}[htbp]
\centering
\epsfig{file=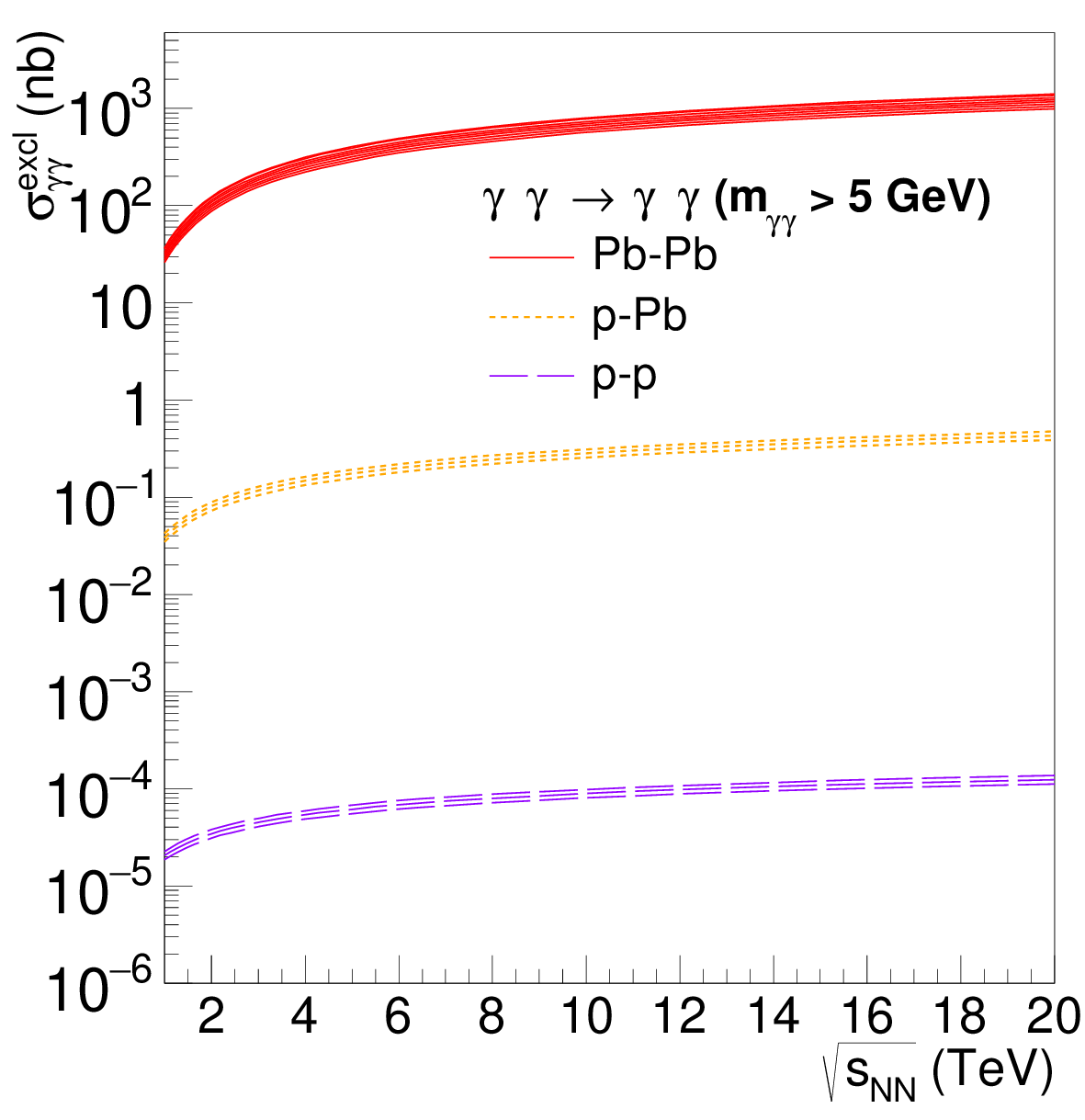,width=0.49\columnwidth,height=8.cm}
\epsfig{file=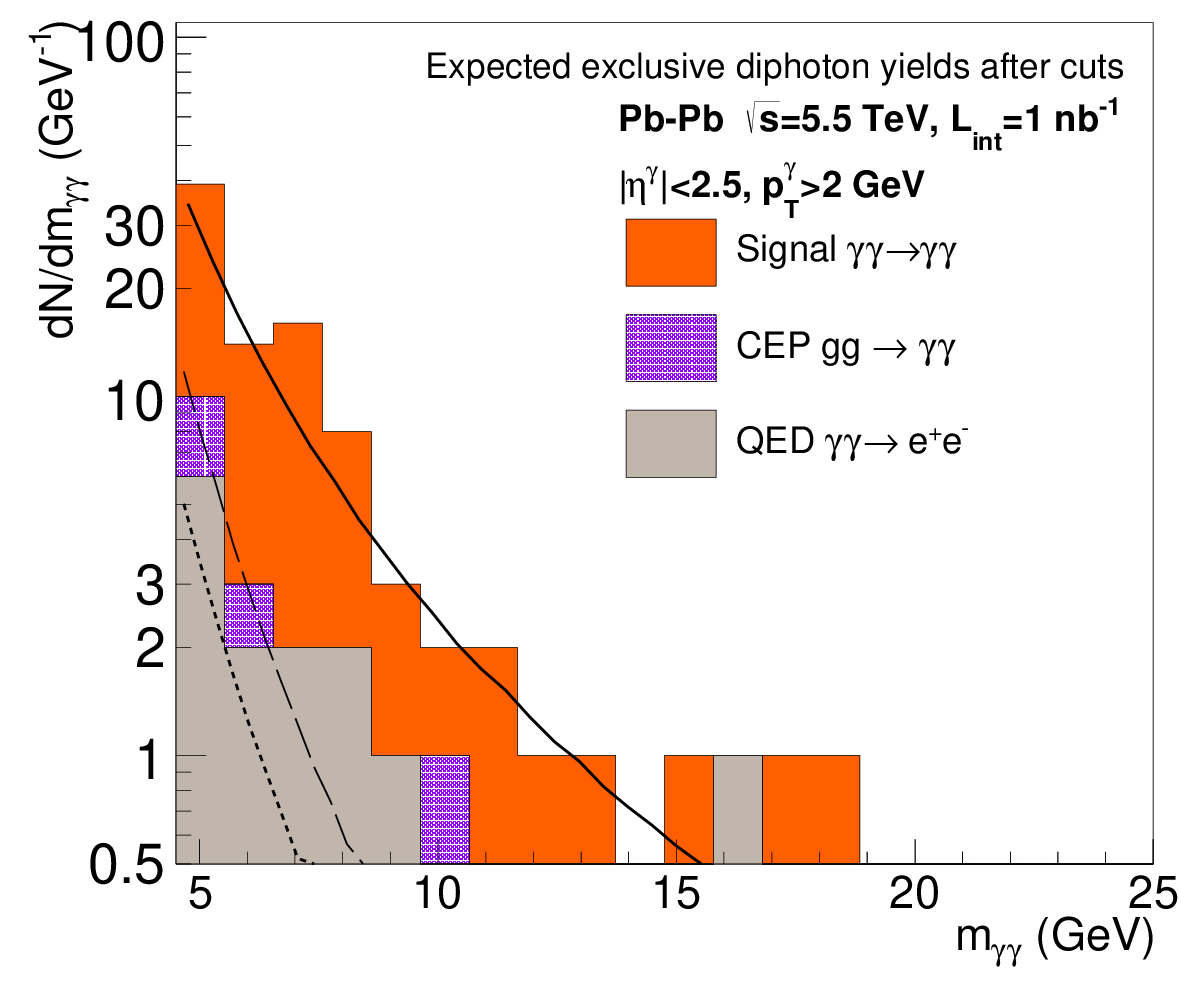,width=0.49\columnwidth,height=8.cm}
\caption{Left: Cross sections for exclusive $\gaga\to\gaga$, with pair masses above 5~GeV, in ultraperipheral \PbPb\
(top curve), \pPb\ (middle) and \pp\ (bottom) collisions as a function of the nucleon-nucleon c.m.{} energy in the
range $\sqrtsnn$~=~1--20~TeV. 
Right: Stacked diphoton yields as a function of invariant mass for elastic $\gaga$ and backgrounds (CEP $\gaga$ and
QED e$^+$e$^-$) expected in 1~nb$^{-1}$ \PbPb\ collisions at $\sqrtsnn$~=~5.5~TeV after analysis cuts 
(see text). The three superimposed curves indicate the underlying individual LbyL (solid), QED (dashed)
and CEP (dotted) distributions after cuts.}
\label{fig:1}
\end{figure}

Our $\gaga\to\gaga$ calculations are carried out for a minimum diphoton mass $\mgg$~=~5~GeV.
Such a choice is driven by three considerations. First, the final state lies in the continuum region
above the range where contributions from two-photon decays from exclusively produced hadronic resonances 
dominate --the $p$-wave scalar and tensor charmonium states 
$\chi_{\rm c0,c2}$ at masses 3.4--3.9~GeV are the heaviest particles with an observed $\gaga$ decay~\cite{PDG} 
before the Higgs boson (the $\chi_{\rm b0,b2}$ at 
around 10~GeV should have also a diphoton decay but it has not been observed so far).
Second, experimentally one needs a signal of a few GeV in the calorimeters in order to
reliable trigger the acquisition of the event above noise and avoiding exclusive final-states
with softer photons from decays of lower-mass hadrons ($\pi^0$, $\eta$, K$^0_s$, ...) with much larger
cross sections. Third, the $\gaga$ cross section for diphoton masses below 
5~GeV has larger theoretical uncertainties 
as the hadronic LbyL contributions are computed less reliably by the quark boxes~\cite{Bern:2001dg}.
Using the theoretical setup described in the previous section we obtain the values of 
$\sigmagg^{\rm excl}$[$\mgg>$~5~GeV] at the LHC listed in Table~\ref{tab:1}. 
In Fig.~\ref{fig:1} (left) we show the predictions for the three systems 
in a wider range of c.m.{} energies, $\sqrtsnn$~=~1--20~TeV. The cross sections are in the 
hundreds of fb, pb, and nb for \pp, \pPb, and \PbPb\ respectively, clearly showing the importance of
the Z$^2$ single photon-flux enhancement factor for ions compared to protons.
Our \pp\ result at 14~TeV 
is consistent with the one obtained in~\cite{Khoze:2004ak}, 
whereas those for \pPb\ and \PbPb\ are calculated in this work for the first time.\\


Realistic estimates of the detectable number of $\gaga\to\gaga$ events at the LHC can be obtained by considering
the luminosity for each colliding system, geometric acceptance of the detectors, experimental 
efficiencies, possible instrumental biases, and potential backgrounds.
We focus on the ATLAS~\cite{Aad:2008zzm} and CMS~\cite{Chatrchyan:2008aa} experiments which feature photon
detection capabilities with tracking and calorimetry over 5 units of pseudorapidity
($|\eta\,|<$~2.5), plus forward detectors (up to at least $|\eta\,|$~=~5) needed to tag 
rapidity-gaps on both sides of the central diphoton system, 
and zero-degree calorimeters (ZDC) to veto very-forward-going neutral fragments in 
ion collisions which help to further reduce backgrounds (see below). 
One cannot unfortunately use the existing TOTEM~\cite{totem} and ALFA~\cite{alfa} Roman Pots
to tag the electromagnetically-scattered protons because their acceptance 
below $\mgg\approx$~200~GeV is very small\footnote{Proposed future proton spectrometers at $\pm$420~m~\cite{fp420}
  have a better acceptance for the lower diphoton masses of interest here.}.  
The narrower single-photon acceptances of ALICE ($|\eta\,|\lesssim$~0.9) 
and LHCb ($\eta\approx$~2--5) would reduce the visible diphoton rates by at least a factor of four.\\

In order to carry out the measurement one needs beam luminosities which minimize the number of simultaneous collisions per
bunch-crossing (``pileup'') so as to keep the rapidity gaps in both hemispheres adjacent to the central diphoton
system free of hadronic activity from overlapping collisions. The \pPb\ and \PbPb\ luminosities 
are low enough to keep the pileup well below one, and
one can take their full expected integrated luminosity per run 
(Table~\ref{tab:1}) as usable for the measurement. In the \pp\ case, the average pileup
is as high as 30 and, thus, we will indicatively consider that only 1\% of the nominal 100~fb$^{-1}$/year
can be collected under low pileup conditions.
In such scenarios, one can easily record $\gaga\to\gaga$ events with a
$\varepsilon_{\rm trig}\approx$~100\% efficient trigger based on two back-to-back photons in the
central detector with transverse momentum $\rm p_{T}^\gamma>$~2~GeV plus 
a large rapidity gap $\Delta\eta\gtrsim$~2.5, 
as done in~\cite{Chatrchyan:2012tv}.
We use the \madgraph\ MC to generate elastic $\gaga$ scattering events and simulate the effect of
the ATLAS and CMS geometrical acceptance. The requirement to have both photons with $|\eta^\gamma|~<$~2.5 and
$\rm p_{T}^\gamma>$~2~GeV, reduces the yield by $\varepsilon_{\rm acc}\approx$~0.2 in \pp\ and \pPb\
collisions, and by $\varepsilon_{\rm  acc}\approx$~0.3 in the \PbPb\ case where the photon fluxes are softer
and the diphoton system is produced at more central rapidities. 
We further consider typical offline photon reconstruction and identification efficiencies of order
$\varepsilon_{\rm rec,id\,\gamma}\approx$~80\% in the energy range of interest~\cite{Chatrchyan:2012tv}.  
The final combined signal efficiency is 
$\varepsilon_{\rm pp,pPb\to\gaga}=\varepsilon_{\rm trig}\cdot\varepsilon_{\rm acc}\cdot\varepsilon_{\rm
  rec,id\,\gamma}^2\approx$~0.12 for \pp\ and \pPb\ and $\varepsilon_{\rm PbPb\to\gaga}\approx$~0.20 for \PbPb.
The number of events expected per year in ultraperipheral collisions at the LHC
obtained from the product 
$N_{\gaga}^{\rm excl}=\varepsilon_{\gaga}\cdot\sigma_{\gaga}^{\rm excl}\cdot\Lumi_{\rm AB}\cdot\Delta t$,
are listed in Table~\ref{tab:1}. 
Clearly, the \PbPb\ system provides the best signal counting rates, with associated statistical uncertainties
of order $\pm(\rm N_{\gaga}^{\rm excl})^{1/2}$, i.e. about $\pm$12\%.
The expected diphoton mass distribution for the elastic $\gaga\to\gaga$ signal, 
taking into account acceptance and efficiency losses, normalized by the expected integrated
luminosity in one \PbPb\ run is shown in Fig.~\ref{fig:1} (right) 
compared to the two main residual backgrounds (see below).\\


The central exclusive $\gaga$ measurements~\cite{Aaltonen:2011hi,Chatrchyan:2012tv} confirm that
standard offline event selection criteria:
(i) just two isolated photons within $|\eta\,|<$~2.5 with reconstructed invariant mass $\mgg>$~5~GeV,
(ii) no other charged-particle activity associated with the interaction vertex, 
plus (iii) no other hadronic activity in the event above detector noise over $|\eta|<$~5;
reduce to a negligible level any hadronic interaction except CEP and diffractive Pomeron-induced 
($\Pom\Pom$, or $\gamma\Pom$) 
final-states containing two photons plus rapidity gaps. 
As a matter of fact, for the \pp\ case in the range of $\mgg$ considered here, the LbyL signal
is swamped by the CEP $gg\to\gaga$ cross section which scales with the fourth power of the gluon density
and is 2--3 orders of magnitude larger than the former~\cite{Khoze:2004ak}. 
Using the \superchic~(version 1.41) MC~\cite{superchic}, we obtain a cross section after
acceptance cuts of $\sigma_{gg\to\gaga}^{\rm CEP}$~=~20$^{\times 3}_{\times 1/3}$~pb, where the uncertainties
include the choice of the parton distribution function (PDF) and $\hat{S}^2$ survival factor.
Exclusive $\pi^0\pi^0$ or $\eta(')\eta(')$ production,
decaying into multi-photon final-states, 
can also be a potential background to the diphoton signal. 
These processes have cross sections $\mathcal{O}$(1--100~pb) but
taking into account their $\gamma$ branching ratios and acceptance plus $\mgg$ cuts results in a
negligible final contribution compared to CEP $\gaga$~\cite{Harland-Lang:2013ncy}.\\

Various features can be used to separate $\gaga$-fusion from CEP and, in general, $\Pom$-mediated events. 
Whereas systems from quasi-real photon-fusion are produced almost at rest 
and thus the final-state photons are emitted back-to-back with balanced
pair transverse momentum (p$_{\rm T}^{\gaga}\approx$~0, smeared by the experimental resolution), typical 
CEP photon pairs peak instead at p$_{\rm T}^{\gaga}\approx$~0.5~GeV and have moderate tails
in their azimuthal acoplanarity $\Delta\phi_{\gaga}$. 
Central exclusive $\gaga$ production is the least reducible of all potential backgrounds as other diffractive and
$\gamma$-induced final-states with photons have larger p$_{\rm T}^{\gaga}$ and diphoton acoplanarities. 
By imposing very tight cuts in the pair momentum, p$_{\rm T}^{\gaga}\lesssim$~0.1~GeV and acoplanarity
$\Delta\phi_{\gaga}-\pi\lesssim$~0.04 in our \madgraph\ and \superchic\ samples, we find that
CEP $\gaga$ can be reduced by a factor of about 90 
with minimum losses of the elastic $\gaga$ signal. However, the final LbyL/CEP ratio after cuts, of order 20, is still
too large to make feasible the LbyL observation with proton beams.
The situation is more advantageous for \pPb\ collisions where LbyL is only about 6 times smaller than CEP, as
obtained scaling by A~=~208 the \pp\ cross section at 8.8~TeV, 
$\sigma_{gg\to\gaga}^{\rm CEP}$~=~16$^{\times 3}_{\times 1/3}$~pb,
multiplied by the square of the Pb gluon shadowing ($R_g^{\rm Pb/p}\approx$~0.7, according to the
EPS09 nuclear PDF~\cite{eps09} in the relevant range of gluon fractional momenta
$x\approx$~5$\cdot 10^{-4}$ and virtualities Q$^{2}\approx$~5~GeV$^{2}$). A final LbyL/CEP ratio of order one
is reachable applying the aforementioned p$_{\rm T}^{\gaga}$ and $\Delta\phi_{\gaga}$ cuts. Yet, given the 
low \pPb\ event rates expected (Table~\ref{tab:1}) a potential observation of LbyL scattering
would require an increment of the integrated luminosity (which can be achieved
increasing the p-beam intensity from its conservative default value~\cite{d'Enterria:2009er}) 
and a careful control of the CEP background.\\

In terms of backgrounds, the situation is much more favorable in the \PbPb\ case where hard 
parton-mediated 
exclusive or diffractive cross sections (which scale as A$^2$ compared to \pp) play a
comparatively much smaller role than in \pp\ thanks to the Z$^{4}$-enhancement of electromagnetic interactions. 
In addition, since the nucleus is a fragile object --the nucleon binding energy is just 8~MeV-- even
the softest CEP or \Pom-mediated interactions will result in the emission of a few nucleons from the ion,
detectable in the ZDCs. Thus, studying the activity in the ZDCs can additionally help reduce any residual
diffractive background. The \PbPb\ CEP cross section --as obtained by  A$^2$-scaling the 
$\sigma_{gg\to\gaga}^{\rm CEP}$~=~13$^{\times 2.5}_{\times 0.4}$~pb cross section in \pp\ at 5.5~TeV times the
fourth power of the Pb gluon shadowing-- is 
of the same order as $\sigmagg^{\rm excl}$. Adding a simple p$_{\rm T}^{\gaga}<$~0.2~GeV condition, reduces the CEP
background by $\sim$95\% without any significant loss of signal events.
Other electromagnetic processes similarly enhanced by the Z$^4$ factor
can notwithstanding constitute a potential concern 
if the final-state particles are misidentified as photons. 
Exclusive $\gaga\to\,$e$^+$e$^-$ events\footnote{Fake diphoton signals from other QED processes such as
$\gaga\to\mu^+\mu^-,\tau^+\tau^-,\qqbar$ are much smaller as their final-states include charged particles in
the tracker and/or muon spectrometer.} 
can be misidentified 
if neither electron track is reconstructed 
or if both electrons undergo hard bremsstrahlung. Experimental studies indicate single-electron
misidentification probabilities as low as $f_{\rm e\to\gamma}\approx$~0.5\%~\cite{fakephot},
which can be experimentally confirmed by imposing increasingly stringent photon identification cuts 
and observing the disappearance of the fake diphoton peak from exclusive $\Upsilon\to$e$^+$e$^-$
photoproduction~\cite{cms_hi_ptdr}. Thus, the very large QED cross section in \PbPb, 
$\sigma_{\rm \gaga\to e+e-}^{\rm excl}$[m$_{\rm ee}>$~5~GeV]~$=$~5.4~mb
according to \textsc{Starlight}~\cite{Nystrand:2004vn} --reduced first by a factor of
10 when requiring both e$^+$ and e$^-$ within the central acceptance~\cite{cms_hi_ptdr} and secondly by 
the extra $f_{\rm e\to\gamma}^2$ factor-- 
results in a residual e$^+$e$^-$ contamination of the order of 20\% of the visible
LbyL cross section. Additional (e.g. acollinearity) cuts~\cite{d'Enterria:2009er} 
could be applied to remove any remaining QED di-fermion continuum (notably $\gaga\to\qqbar\to\pi^0\pi^0$) with very
small signal loss. Figure~\ref{fig:1} (right) shows the dominant QED and CEP
backgrounds, expected after cuts in one \PbPb\ run, compared to the LbyL signal as a function of the diphoton
mass. Both contaminations are softer than the signal and their total sum does not exceed the LbyL yields.\\


\paragraph*{Summary.  --}
Despite its fundamental simplicity, no direct experimental observation of light-by-light scattering exists so far.
We have shown that elastic photon-photon scattering can be potentially observed at the LHC using the 
large (quasi-real) photon fluxes in electromagnetic interactions of protons and ions accelerated at TeV energies.
The  $\gaga\to\gaga$ cross sections for diphoton masses in the continuum range above
$\mgg =$~5~GeV 
are 105~fb for \pp, 260~pb for \pPb, and 370~nb for \PbPb\ at the nominal c.m.{} energies, 
clearly showing the importance of the Z$^4$ enhancement of the photon fluxes in ion-ion collisions.
The  number of 
exclusive $\gaga\to\gaga$ events expected in ATLAS and CMS have been obtained taking into account realistic integrated
luminosities and $\gamma$ acceptance and efficiency cuts. 
In the \pp\ case, the dominant background due
to exclusive gluon-induced  production can be reduced imposing cuts on the pair
p$_{\rm T}$ and acoplanarity but unfortunately not to a level where the 
signal can be observed. The signal/background ratio is better in the \pPb\ case but the small expected number of
events 
makes the measurement of the light-by-light signal challenging without (reachable) luminosity increases. 
An unambiguous observation of the process is 
possible in \PbPb\ collisions which provide
$\rm N_{\gaga}^{\rm excl}\approx$~70 elastic photon pairs per run after cuts, with controllable backgrounds. 
The unique measurement of elastic $\gaga$ scattering at the LHC will not only constitute the first experimental
observation of a fundamental quantum mechanical process but may be sensitive to new-physics effects
predicted in various extensions of the SM.\\ 


\paragraph*{Acknowledgments}
\noindent
We thank Diogo Franzoni for support with \madgraph, Lucian Harland-Lang for valuable discussions and feedback
on central-exclusive production in \superchic\ and for $\gaga\to\gaga$ cross sections cross-checks with
\superchic\ 2.0, and Jonathan Hollar for suggestions on the letter.
G.G.S. acknowledges support from CNPq/Brazil.



\end{document}